# From Attention to Citation, What and How Does Altmetrics Work?

Xianwen Wang[a,b]*, Chen Liu[a,b], Zhichao Fang[a,b], Wenli Mao[a,b]

[a]WISE Lab, Faculty of Humanities and Social Sciences, Dalian University of Technology, Dalian 116085, China.
[b]School of Public Administration and Law, Dalian University of Technology, Dalian 116085, China.

* Corresponding author.
Email address: xianwenwang@dlut.edu.cn

**Abstract:** Scholarly and social impacts of scientific publications could be measured by various metrics. In this study, the relationship between various metrics of 63,805 PLOS research articles are studied. Generally, article views correlate well with citation, however, different types of article view have different levels of correlation with citation, when pdf download correlates the citation most significantly. It's necessary for publishers and journals to provide detailed and comprehensive article metrics. Although the low correlation between social attention and citation is confirmed by this study and previous studies, more than ever, we find that social attention is highly correlated with article view, especially the browser html view. Social attention is the important source that bringing network traffic to browser html view and may lead to citation subsequently. High altmetric score has the potential role in promoting the long-term academic impact of articles, when a conceptual model is proposed to interpret the conversion from social attention to article view, and to citation finally.

## 1. Introduction

Citation has been the widely accepted metrics of impact of a scientific publication for decades. In the recent years, many indicators have been introduced to the family of article-level metrics. Besides citations, the impact of an article could be reflected and quantified by article views, readerships and Altmetric score, etc. Altmetrcis (Priem & Costello, 2010; Priem, Costello, & Dzuba, 2011) tracks the online mentions by pulling in data from social media, blog, traditional media and online reference managers. Recent efforts have been made to investigate the effect of altmetrcis on scholarly impact. From the qualitative prospect, Sud & Thelwall draw an abstract diagram of factors affecting the relationship between altmetrics and citation counts (Sud & Thelwall, 2014). Shuai et al. use the dataset of 4,606 arXiv preprint submissions during a period of 7 months to study the online response to new articles, they find that the volume of Twitter mentions is statistically correlated with arXiv downloads, in addition, early citations may favor highly mentioned articles (Shuai, Pepe, & Bollen, 2012). There are statistically significant associations between higher metric scores and higher citations for articles with positive altmetric scores (Thelwall, Haustein, Larivière, & Sugimoto, 2013). Eysenbach's findings also confirm that

Tweets can predict highly cited articles in only 3 days after the article publication (Eysenbach, 2011). However, Haustein et al. show that the correlation between citation and tweets is relatively low (Haustein, Peters, Sugimoto, Thelwall, & Larivière, 2014). Li et al. report that the correlation between online readership and traditional citations count is significant (Li, Thelwall, & Giustini, 2012).

Nowadays, many STM publishers and journals individually provide the article usage data to evaluate the impact of an individual article. With this kind of data, it is possible for researchers to explore the online usage of scientific publications, e.g., studying the researchers' working timetable through the downloaded time of articles (Wang, et al., 2012; Wang, et al., 2013), tracing the research trends combining the downloaded time and downloaded content (Wang, Wang, & Xu, 2013), exploring the dynamic patterns of downloaded time (Wang, Mao, Xu, & Zhang, 2014; Wang, Wang, & Mao, 2014). For the relationship between article downloads and citations, Watson shows that compared to downloads, citations are delayed by about 2 years, so download statistics provide a useful indicator of eventual citations in advance (Watson, 2009). In a previous analysis, Lippi et al. conclude that there is a strong relationship between article downloads and citations, the most downloaded articles are also those that are more likely to receive citations (Lippi & Favaloro, 2012). Jahandideh et al. also find that more downloads at a limited period of time is an indicator of more citations to the article in a long-term interval (Jahandideh, Abdolmaleki, & Asadabadi, 2007). O'Leary conducts an analysis of the relationship between the numbers of downloads and citations by examining the most heavily downloaded articles published in *Decision Support Systems*, he finds that the number of citations and downloads are closely related (O'Leary, 2008). However, Coats's study reveals that citations and downloads show diverse preferences, so we need to keep track of both types of impact (Coats, 2005).

Significant correlation between views and citations is confirmed by many previous studies, however, the accuracy of distinct measurements should be examined separately because each of the view patterns has its individual characteristics. Yan and Gerstein find that although spreading patterns of scientific information are remarkably similar in data from different journals, there are intrinsic differences for different types of article usage (HTML views and PDF downloads versus XML) (Yan & Gerstein, 2011). PDF downloads increases the probability that people would later read it (Allen, Stanton, Pietro, & Moseley, 2013). Our research questions are that, do different types of view patterns make same impact on citations? Furthermore, does distinguishing different types of view patterns offer a more reasonable description for measuring the impact of an article? Although previous studies show that there is low correlation between social media and citation, how is the relationship between social media and other metrics, e.g., article view?

## 2. Data

PLOS publications are selected as our research objects. The reasons are, firstly, articles published by PLOS are all open-access, which could eliminate the bias between open-access and non open-access data; secondly, PLOS provide detailed article-level metrics data for each individual publication. PLOS article view data are provided in three different formats, which are HTML, PDF, and XML, when the online activities of users across these three formats are recorded and reported under COUNTER 3 standards.

The article metrics data in this study is harvested directly from the PLOS Article-Level-Metrics platform (http://almreports.plos.org), which allows researchers to view article-level metrics for any set of PLOS articles. The altmetric data is collected from www.Altmetric.com.

PLOS publishes 7 journals, as Table 1 shows, in which PLOS ONE is a multidisciplinary journal and has the most articles, the number is 52,258, much more than other field-oriented journals.

**Table 1** PLOS journals list

| Journal | Abrev | No. of articles in the dataset | Started year | IF |
|---|---|---|---|---|
| PLOS ONE | plosone | 52,258 | 2004 | 3.730 |
| PLOS Genetics | plosgenetics | 2,880 | 2005 | 8.517 |
| PLOS Pathogens | plospathogens | 2,562 | 2005 | 8.136 |
| PLOS Computational Biology | ploscompbiol | 2,231 | 2005 | 4.867 |
| PLOS Biology | plosbiology | 1,612 | 2003 | 12.690 |
| PLOS Neglected Tropical Diseases | plosntds | 1,504 | 2007 | 4.569 |
| PLOS Medicine | plosmedicine | 859 | 2004 | 15.253 |

In this study, only the research articles are kept in the dataset, when other article types are excluded. The metrics selected include article view, Mendeley readership, altmetric score. There are several types of article view, which are PLOS browser html view, PLOS pdf view/download, PLOS total view, PMC browser html view, PMC pdf view/download and PMC total view.

**3. Results**
**3.1 Descriptive Statistics**
64,305 PLOS research articles are divided by the publication years. Besides, data is classified into two periods, which are the older period when articles are published more than 5 years and without altmetric data (2004-2008, when the social web is in its initial developing) and the young period when articles are published in the recent 5 years with Altmetric data parsed (2009-2012). For example, 7709 article published in the period of 2004-2008 have averagely 8402 views, and 45 readers in Mendeley, when the 56,196 articles during the period of 2009-2012 averagely have 3854 views, 21 Mendeley readers and Altmetric scores of 3.

For the annual data from 2004 to 2012, the annual descriptive statistics of total article view (including PLOS total and PMC total), Mendeley readership and Altmetric score are given in Table 2. The orange data bars in the table can spot larger and smaller numbers and demonstrate the trend clearly. A longer bar represents a larger value, and vice versa. We also provide a kind of tiny chart, namely line spakline, as another visual representation to show the yearly trends for the statistics of the annual data (data column). The line sparklines are placed in the cells in the row of 2004-2012.

As the data bars and line sparklines display, compared with the downward trend of the total usage and mendeley, the Altmetric score shows an upward trend. For example, the average article view descends from 16,667 for articles in 2004 to 3025 for articles in 2012. And the statistic for Mendeley readership shows the same trend. However, for the Altmetric score, 6038 articles in 2009 get the score of 1.95 averagely; for 25,634 articles published in 2012, the average Altmetric score is as high as 3.89. About 31.37% of all articles published in 2009-2012 have Altmetric score greater than 0.

**Table 2** Descriptive statistics of PLOS articles divided by year

| Year | Number | Scopus citation | | Article views | | Mendeley readership | | Altmetrics score | |
|---|---|---|---|---|---|---|---|---|---|
| | | Mean | Median | Mean | Median | Mean | Median | Mean | Median |
| 2004 | 189 | 135.09 | 97 | 16,667 | 12,253 | 88 | 54 | | |
| 2005 | 419 | 73.79 | 43 | 13,199 | 9,944 | 72 | 50 | | |
| 2006 | 899 | 51.36 | 33 | 10,836 | 8,401 | 56 | 37 | | |
| 2007 | 2,199 | 40.18 | 26 | 8,097 | 6,074 | 44 | 29 | | |
| 2008 | 4,403 | 37.03 | 24 | 7,131 | 5,249 | 39 | 25 | | |
| 2009 | 6,038 | 27.55 | 19 | 5,873 | 4,436 | 33 | 22 | 1.95 | 0 |
| 2010 | 8,653 | 18.40 | 13 | 4,932 | 3,756 | 30 | 20 | 2.31 | 0 |
| 2011 | 15,871 | 11.59 | 8 | 3,837 | 2,970 | 22 | 15 | 2.68 | 0 |
| 2012 | 25,634 | 5.31 | 4 | 3,025 | 2,212 | 14 | 10 | 3.89 | 1 |
| 2004-2012 | | | | | | | | | |
| 2004-2008 | 7,709 | 44 | 27 | 8,402 | 6,127 | 45 | 29 | | |
| 2009-2012 | 56,196 | 11 | 7 | 3,854 | 2,837 | 21 | 13 | 3 | 0 |

As shown in Table 3, the whole dataset is also divided by seven journals and two periods. For the 2004-2008 period, articles of *PLOS Medicine* has the most average citation and article view, when articles of *PLOS Biology* has the most Mendeley readership, the second most citation and the almost same article view as *PLOS Medicine*. Overall, as shown by the orange data bas in Table 3, the trends of the average article view coincides well with the average citation. However, the trend of data bars in the Mendeley readership columns are rather different with the previous ones. For example, PLOS Medicine has the most average citations and article views, but doesn't have many Mendeley readers; when PLOS Computational Biology has the most Mendeley readers, but the below average citations and article views.

For the 2009-2012 period, citation and article view still have the same trends, but very different with the trends of the Mendeley readership and Altmetric score.

**Table 3** Descriptive statistics of PLOS articles divided by journals and periods

| Dateset | Number | Scopus citation | | Article views | | Mendeley readership | | Altmetrics score | |
|---|---|---|---|---|---|---|---|---|---|
| | | Mean | Median | Mean | Median | Mean | Median | Mean | Median |
| plosmedicine 04-08 | 489 | 96.05 | 63 | 14,907 | 10,258 | 44 | 31 | | |
| plosbiology 04-08 | 938 | 78.15 | 48 | 14,105 | 10,517 | 97 | 68 | | |
| plospathogens 04-08 | 577 | 46.54 | 35 | 7,809 | 6,654 | 34 | 27 | | |
| plosgenetics 04-08 | 774 | 44.24 | 28 | 9,221 | 7,138 | 55 | 38 | | |
| ploscompbiol 04-08 | 645 | 34.05 | 24 | 7,441 | 6,301 | 60 | 46 | | |
| plosone 04-08 | 4,128 | 31.92 | 22 | 6,494 | 4,767 | 32 | 21 | | |
| plosntds 04-08 | 159 | 26.16 | 19 | 6,416 | 5,500 | 22 | 18 | | |
| plosmedicine 09-12 | 370 | 37.15 | 22 | 13,015 | 9,808 | 39 | 31 | 19.59 | 6 |
| plosbiology 09-12 | 674 | 36.09 | 25 | 10,382 | 8,360 | 71 | 51 | 8.91 | 2 |
| plospathogens 09-12 | 1,985 | 23.93 | 18 | 5,667 | 4,930 | 27 | 22 | 2.57 | 0 |
| plosgenetics 09-12 | 2,106 | 24.67 | 16 | 6,575 | 5,390 | 42 | 30 | 3.35 | 1 |
| ploscompbiol 09-12 | 1,586 | 15.92 | 10 | 5,507 | 4,442 | 49 | 36 | 2.67 | 0 |
| plosone 09-12 | 48,130 | 9.68 | 6 | 3,440 | 2,576 | 18 | 12 | 2.94 | 0 |
| plosntds 09-12 | 1,345 | 12.59 | 9 | 4,330 | 3,666 | 20 | 15 | 2.18 | 0 |

### 3.2 Correlation analysis

*Correlation between citation and other metrics*

Usually, Spearman instead of Pearson correlation is used because metrics data is typically too skewed for the assumption of normal distribution of a Pearson test and has too many zero values to be transformed into a normal distribution (Li, Thelwall, & Giustini, 2012; Sud & Thelwall, 2014). As shown in Table 4, the annual Spearman correlation coefficients between Scopus citation and other metrics are calculated. Other metrics include the Total article view (combined by PLOS total and PMC total), HTML view (combined by PLOS total and PMC total), PDF view (combined by PLOS total and PMC total), Mendeley readership and Altmetric score, which are abbreviated with the first letter of T, H, P, M, A, as the last column shows in Table 4.

To better demonstrate the result, here we use color scales and column sparklines to show the show the general distribution of values. For color scales, cells are shaded with gradations of colors that correspond to minimum (green), midpoint (yellow), and maximum values (red). Besides, the column sparklines in the last column show the performance of the values, and highlight the highest value.

As Table 4 shows, all correlation coefficients are statistically significant at the 1% level. For the three kinds of view, the correlation between Scopus citation and HTML view is the weakest, when the correlation between citation and PDF view are stronger than others. In most sub dataset divided by publication year, the correlation coefficients are higher than 0.650. It seems that PDF view has the greater potentiality than HTML browser view to lead to academic citation.

The correlations between citation and Mendeley readership are around 0.500, which is much lower than the view, especially the values in the PDF column. And the correlation between citation and Altmetric score is very weak, when the coefficients ranges from 0.14 to 0.22.

**Table 4** correlation between citation and other metrics of articles divided by years

| Dateset | N | Total | HTML | PDF | Mendeley | Altmetrics | T H P M A |
|---|---|---|---|---|---|---|---|
| 2004 | 189 | 0.705 | 0.683 | 0.772 | 0.500 | NA | ■■■■ |
| 2005 | 419 | 0.515 | 0.503 | 0.517 | 0.363 | NA | ■■■■ |
| 2006 | 899 | 0.492 | 0.441 | 0.629 | 0.469 | NA | ■■■■ |
| 2007 | 2199 | 0.581 | 0.549 | 0.672 | 0.525 | NA | ■■■■ |
| 2008 | 4003 | 0.676 | 0.648 | 0.740 | 0.550 | NA | ■■■■ |
| 2009 | 6038 | 0.664 | 0.637 | 0.719 | 0.547 | 0.221 | ■■■■_ |
| 2010 | 8653 | 0.612 | 0.587 | 0.658 | 0.484 | 0.198 | ■■■■_ |
| 2011 | 15871 | 0.627 | 0.606 | 0.651 | 0.490 | 0.143 | ■■■■_ |
| 2012 | 25634 | 0.569 | 0.544 | 0.608 | 0.420 | 0.149 | ■■■■_ |
| 2004-2008 | 7709 | 0.631 | 0.603 | 0.699 | 0.538 | NA | ■■■■ |
| 2009-2012 | 56196 | 0.673 | 0.647 | 0.715 | 0.546 | -0.031 | ■■■■_ |

Note: All correlation coefficients in Table 4 are statistically significant at the 1% level.

To find that whether there exist difference among different journals, the dataset is divided by the seven journals and two periods. As Table 5 shows, the correlations between citation and PDF view/download are higher than the coefficients between citation and other metrics, when the values in the column of HTML are lower than columns of PDF and Total. And moreover, the correlation between citations and Altmetric scores is slight negative.

**Table 5** correlation between citation and other metrics of different journals divided by two periods

| Dateset | N | Total | HTML | PDF | Mendeley | Altmetrics | T H P M A |
|---|---|---|---|---|---|---|---|
| plosone 04-08 | 4128 | 0.600 | 0.563 | 0.715 | 0.530 | NA | ■■■■ |
| plosone 09-12 | 48130 | 0.654 | 0.627 | 0.694 | 0.506 | -0.058 | ■■■■_ |
| plosgenetics 04-08 | 774 | 0.610 | 0.595 | 0.625 | 0.540 | NA | ■■■■ |
| plosgenetics 09-12 | 2106 | 0.642 | 0.606 | 0.714 | 0.602 | <u>-0.013</u> | ■■■■_ |
| plospathogens 04-08 | 577 | 0.567 | 0.545 | 0.614 | 0.550 | NA | ■■■■ |
| plospathogens 09-12 | 1985 | 0.544 | 0.512 | 0.612 | 0.531 | -0.073 | ■■■■ |
| ploscompbiol 04-08 | 645 | 0.503 | 0.480 | 0.549 | 0.543 | NA | ■■■■ |
| ploscompbiol 09-12 | 1586 | 0.584 | 0.552 | 0.670 | 0.546 | -0.095 | ■■■■_ |
| plosbiology 04-08 | 937 | 0.598 | 0.592 | 0.565 | 0.459 | NA | ■■■■ |
| plosbiology 09-12 | 674 | 0.500 | 0.458 | 0.640 | 0.499 | <u>-0.033</u> | ■■■■_ |
| plosntds 04-08 | 159 | 0.400 | 0.378 | 0.489 | 0.399 | NA | ■■■■ |
| plosntds 09-12 | 1345 | 0.406 | 0.322 | 0.635 | 0.528 | -0.141 | ■■■■_ |
| plosmedicine 04-08 | 489 | 0.582 | 0.549 | 0.662 | 0.586 | NA | ■■■■ |
| plosmedicine 09-12 | 370 | 0.467 | 0.405 | 0.689 | 0.611 | *-0.122* | ■■■■_ |

Note: All correlation coefficients in Table 5 except the two cells with underlines and one cell with italic are statistically significant at the 1% level. Cell with italic is statistically significant at the 5% level.

*Correlation between Altmetric score and other metrics*

Although the correlation between Altmetric scores and citations are weak, how about the relationship between Altmetric scores and other metrics? Table 6 shows that the correlation coefficients between Altmetric scores and other metrics are much higher than the values in the column of citation.

For the coefficients between Altmetric score and three metrics of view (Total, HTML and PDF), the column of HTML has the highest correlations (range from 0.35 to 0.40), when the column of PDF is the lowest (range from 0.23 to 0.30).

**Table 6** correlation between Altmetric score and other metrics divided by years

| Year | N | Citation | Total | HTML | PDF | Mendeley | C T H P M |
|------|------|----------|-------|-------|-------|----------|-----------|
| 2009 | 6038 | 0.221 | 0.383 | 0.395 | 0.296 | 0.339 | |
| 2010 | 8653 | 0.198 | 0.359 | 0.371 | 0.273 | 0.336 | |
| 2011 | 15871 | 0.143 | 0.344 | 0.353 | 0.233 | 0.29 | |
| 2012 | 25634 | 0.149 | 0.379 | 0.4 | 0.236 | 0.342 | |

Note: All correlation coefficients in Table 6 are statistically significant at the 1% level.

Then, we calculate the annual correlation coefficients between Altmetric score and citation and three kinds of HTML views (PLOS HTML views, PMC HTML views, and PLOS and PMC HTML view combined) for each journal. As Table 7 shows, for each year, the row of PLOS HTML view has the highest correlations, slightly higher than the row of Total HTML view, and much higher than the PMC HTML view.

Among the seven journals, *PLOS Biology* and *PLOS Medicine* have the highest correlation coefficients. For the 103 articles published in *PLOS Medicine* in 2011, the correlation coefficient between Altmetric score and PLOS HTML view is as high as 0.711. And for the 134 articles published in *PLOS Biology* in 2012, the coefficient is 0.733, suggesting a high correlation between Altmetric score and browser view.

**Table 7** correlation between Altmetric score and citations and HTMLview for journals divided by years

| Year | | plos one | plos genetics | plos pathogens | plos compbiol | plos biology | plos ntds | plos medicine |
|---|---|---|---|---|---|---|---|---|
| 2009 | N | 4400 | 436 | 423 | 343 | 181 | 178 | 77 |
| | Citation | 0.179** | 0.317** | 0.136** | 0.281** | 0.159* | 0.06 | 0.221 |
| | PLOS | **0.385**** | **0.399**** | **0.208**** | **0.400**** | **0.538**** | **0.375**** | **0.440**** |
| | PMC | 0.149** | 0.244** | 0.15** | 0.292** | 0.075 | 0.154* | 0.230* |
| | Total | 0.368** | 0.397** | 0.215** | 0.399** | 0.512** | 0.347** | 0.417** |
| 2010 | N | 6734 | 449 | 486 | 392 | 197 | 307 | 88 |
| | Citation | 0.160** | 0.178** | 0.204** | 0.199** | 0.296** | 0.255** | 0.283** |
| | PLOS | **0.359**** | **0.347**** | **0.322**** | **0.316**** | **0.494**** | **0.326**** | **0.572**** |
| | PMC | 0.120** | 0.099* | 0.220** | 0.205** | 0.174* | 0.246** | 0.352** |
| | Total | 0.344** | 0.328** | 0.315** | 0.324** | 0.464** | 0.315** | 0.562** |
| 2011 | N | 13769 | 539 | 518 | 384 | 162 | 396 | 103 |
| | Citation | 0.102** | 0.262** | 0.129** | 0.068 | 0.095 | 0.086 | 0.088 |
| | PLOS | **0.335**** | **0.457**** | **0.276**** | **0.362**** | **0.461**** | **0.286**** | **0.711**** |
| | PMC | 0.071** | 0.217** | 0.154** | 0.085 | 0.125 | 0.156** | 0.232* |
| | Total | 0.314** | 0.446** | 0.266** | 0.339** | 0.449** | 0.282** | 0.677** |
| 2012 | N | 23227 | 682 | 558 | 467 | 134 | 464 | 102 |
| | Citation | 0.112** | 0.291** | 0.155** | 0.147** | 0.232** | 0.114* | 0.226* |
| | PLOS | **0.382**** | **0.509**** | **0.476**** | **0.428**** | **0.733**** | **0.261**** | **0.653**** |
| | PMC | 0.091** | 0.176** | 0.159** | 0.213** | 0.18* | 0.101* | 0.275** |
| | Total | 0.363** | 0.491** | 0.455** | 0.420** | 0.724** | 0.254** | 0.638** |

Note: ** statistically significant at the 1% level, * statistically significant at the 5 level.

Why does the Altmetric score correlate weak with the citation? Firstly, the boom of social media platforms begin in 2009, however, even for the articles published in 2009, there is not enough years for publications to accumulate the citations, let alone for the later articles. Secondly, there is not much possibilities lead social attention directly to academic citation. Research that is popular with the public or interesting even to those outside of the academy tends to have high Altmetric scores. However, many social media participants are not academic researchers, so it's impossible for them to write papers and cite the articles they retweeted. Although there may be many retweets for an article, just a few of them may result in academic citations.

Here we propose a conceptual model to visualize the relationship between social attention, article view and academic citation. As Fig. 1 shows, social attention and academic citation are two ends of the chain. There is little chance for people to cite an article without even reading it. So, there is no link from the end of social attention to another end of citation in Fig. 1. Social attention may bring network traffic to browser html view, and some of the html view may bring about citations directly, when some may lead to pdf downloads and a part of them turn into citations.

Usually, the article links given in social platforms are html rather than pdf links.

Different routes in Fig. 1 are visualized in different colors. The yellow arrows show the routes from social attention to browser html view, to pdf view/download, and then to citation, when the purple arrows demonstrate the routes from browser html view to pdf view/download to citation. The green arrow shows the route from browser to social attention, when the blue arrow is the spread in the social media platforms. The decreased arrow size of the same color indicates the conversions rates from one status to the next.

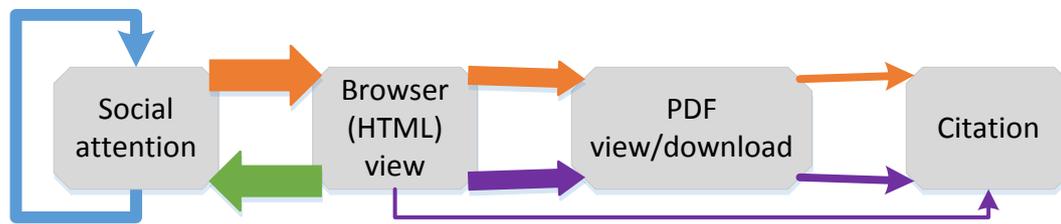

**Fig. 1** Conceptual model of the relationship between social attention, article view and citation

**Discussion**
As more and more academic publishers and journals provide article usage data to public in recent years, scientometrics researchers are embracing the ever-increasing research data. But the data formats of article usage provided by different publishers and journals are lack of the uniform standard. Although some of them like PLOS, Science, PNAS provide the detailed article-level metrics data with different types separately, including browser html view, pdf download, etc. However, for many other publishers and journals which provide article usage statistics, only the total number of article view is available, when the browser html view and pdf download are included in this total number and not differentiated. These publishers include Taylor & Francis, IEEE Xplore Digital Library, Nature, ACM Digital Library.

According to the findings in this study, among different types of article view, pdf download has the most significant correlation with citation. So it is very important for researchers to distinguish different types of article view and necessary for publishers and journals to provide the detailed and comprehensive formats of metrics data. Generally, the types of article view include abstract view, browser html view, and pdf view/download.

Although correlation between social attention and citation are rather low, however, we find that the very significant correlation between altmetric score and article view, especially for browser html view. A conceptual model is proposed to interpret the conversion rate from social attention to article view, and to citation finally. Social attention doesn't have the influence on the academic citation directly, however, it may cause article view and lead to citation subsequently, which means that a part of citation may be caused by social attention, including twitter, facebook, blog, news report, etc.

**Acknowledgements**
The work was supported by the project of "National Natural Science Foundation of China" (61301227).